\documentstyle[aps,prl,multicol,psfig]{revtex} 
\begin{document}
\draft
\title{From Molecular Dynamics to Dissipative Particle 
Dynamics}
\author{Eirik G. Flekk\o y$^1$ and Peter V. Coveney$^2$}
\address{
$^1$Department of Physics, University of Oslo,\\
        P.O. Box 1048 Blindern, 0316 Oslo 3, Norway\\
$^2$  Centre for Computational Science, Queen Mary and 
      Westfield College,\\
      University of London, Mile End Road, London E1 4NS,
 United Kingdom}
\date{\today}
\maketitle 
\begin{abstract}
A procedure is introduced for deriving a coarse-grained 
dissipative particle dynamics from molecular dynamics. 
The rules of the dissipative particle dynamics are derived
from the underlying molecular interactions,
and a Langevin equation is obtained 
that  describes the forces experienced by the dissipative 
particles and specifies the associated canonical Gibbs distribution
for the system.
\end{abstract}

\begin{multicols}{2}
Hydrodynamic simulations of complex fluids remain 
a major challenge in most cases of interest. Such fluids 
include particulate and colloidal suspensions, polymeric liquids, 
emulsions and other self-assembling amphiphilic fluids, and 
fluids where Brownian motion is important.
For such fluids  it is often necessary to base the modeling
on a microscopic picture of the system, 
thus working from the bottom upwards.
Over the last decade several such `bottom up' strategies  have
been introduced.
Hydrodynamic lattice gases \cite{frisch86},
which model the fluid as a discrete set of particles, represent
a computationally efficient discretization of the 
more conventional molecular dynamics (MD)\cite{koplik95}. 

A recent contribution to the family of bottom-up approaches
is the dissipative particle dynamics (DPD) method introduced
by Koelman and Hoogerbrugge in 1992~\cite{hoogerbrugge92}.
Applications of the technique include colloidal
suspensions~\cite{boek96}, polymer solutions~\cite{schlijper95} and
binary immiscible fluids~\cite{coveney96}.
For specific applications where comparison is possible,
this model is orders of magnitude faster than MD~\cite{groot97}.

The basic components of DPD are particles that are 
thought to represent mesoscopic elements of the 
underlying molecular fluid. 
These dissipative particles then evolve just as MD 
particles but with different inter-particle forces: Since 
the DPD particles are pictured as having 
internal degrees of freedom, the forces between them have 
both a fluctuating and a dissipative component in addition
to the conservative forces that are present already at the MD level.
Nevertheless, momentum conservation along with mass conservation produce 
hydrodynamic behavior at the macroscopic level.

Dissipative particle dynamics has been demonstrated to connect
correctly to the 
macroscopic continuum theory; that is, for a one-component DPD
fluid, it is possible  to derive the Navier-Stokes 
equations and to compute the viscosity in the 
large scale limit~\cite{espanol95,marsh98c}.
However, thus far no attempt has been made to link DPD to the underlying 
microscopic dynamics. This is the purpose of the present letter.
We define the dissipative particles (DP) by appropriate 
weight functions that sample a portion  of the underlying 
conservative MD particles, and we derive the 
forces between the DP's from the hydrodynamic 
description of the MD system. 

\begin{figure}
\centerline{\hbox{\psfig{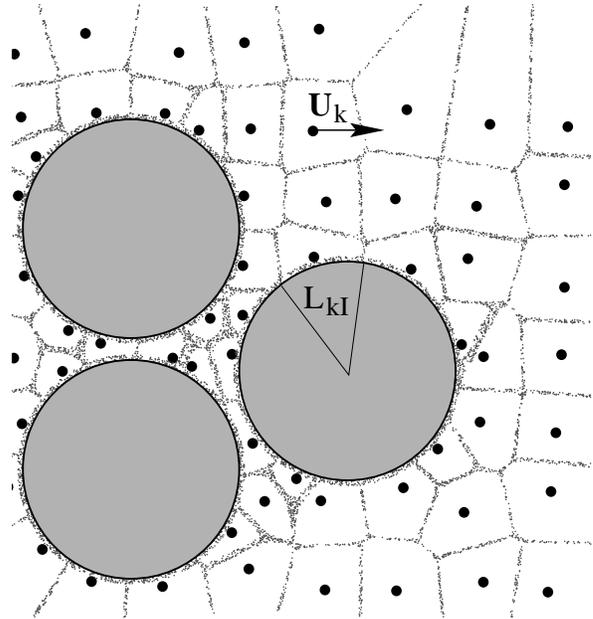}}}
\caption{\label{fig1}
\protect \narrowtext {\bf Multiscale modeling:}
The dissipative particles are defined as cells
in the Voronoi lattice, moving with velocity ${\bf U}_k$. 
There are four relevant length scales:
The scale of the large, gray  colloid particles,
the two scales of the dissipative particles in between and
away from the colloids and finally the scale of the 
MD particles, which are shown as the little dots that form  the 
boundaries between the DP's. }
\end{figure}

The present development has two main virtues,
one fundamental and one practical.
From a fundamental point of view our work gives a 
microscopic foundation to DPD and thus 
provides a quantitative meaning to the term `mesoscopic'.
On the practical 
side this foundation may be used to 
deal with physical systems where the modeling 
is challenged by the simultaneous presence of several
different length scales.
While conventional DP's are spheres of fixed size and mass, the current DP's
are defined as cells on a Voronoi lattice with variable
sizes. This provides us with the freedom to define 
 particle sizes according to the local resolution 
requirements--a particle analog to 
adaptive meshes in finite-element simulations  \cite{hole88}.
The concept is illustrated by the simulation of a colloidal suspension, 
which is shown in
Fig.~\ref{fig1}. 
Here the computational effort is adapted 
to meet the local need for detail of description:
it is larger in narrow regions between the particles than 
in the bulk. Previous DPD simulations have had difficulty with dense
colloidal suspensions precisely because the technique is unable to
handle multiple lengthscale phenomena~\cite{boek96}.
Other complex systems where modeling and simulation frequently involve
several simultaneous 
length scales include polymeric and amphiphilic fluids, particularly
in porous media and restricted geometries~\cite{coveney98}.

The basic ingredient in our derivation of DPD
is an appropriate coarse-graining scheme. The dissipative particles 
are defined as clusters  of MD particles
in such a way that the MD particles are all
represented by the dissipative particles. 
A general way to achieve this is via the sampling 
function
$
f_k({\bf x} )= {s({\bf x} - {\bf r}_k )}/{\sum_l s({\bf x} - {\bf r}_l )}
$.
Here the positions ${\bf r}_k$ and ${\bf r}_l$ define the DP 
centers, which  initially may be distributed arbitrarily in space, 
${\bf x}$ is an arbitrary position and $s({\bf x} )$
is some localized function, which we choose as a Gaussian 
$s({\bf x} ) = \exp{( - x^2/a^2)}
$; the distance $a$ sets the scale of the sampling function.
The mass $M_k$, momentum ${\bf P}_k$ and internal energy $E_k$ of the $k$th DP are then 
defined as
\begin{eqnarray}
M_k &=& \sum_i   f_k({\bf x}_i ) m \nonumber \\
{\bf P}_k &= &M_k {\bf U}_k = \sum_i  f_k({\bf x}_i )m {\bf v}_i  \nonumber \\
\frac{1}{2} M_k U_k^2 + E_k &=& \sum_i f_k({\bf x}_i )
\left(  \frac{1}{2} m v_i^2  + \frac{1}{2} \sum_{j\neq i} V(r_{ij}) \right) \label{DPdef}
\end{eqnarray}
where ${\bf x}_i$ and ${\bf v}_i$ are the position and velocity 
of the $i$'th MD particle, all assumed to have identical masses $m$, 
$ V(r_{ij})$ is the MD interparticle potential, and ${\bf U}_k$ is the velocity 
of the $k$'th DP. 
The kinematic condition $\dot{{\bf r}}_k = {\bf U}_k $ completes the definition of the DPD.
The normalization property 
$\sum_k f_k ({\bf x} ) = 1$ implies directly that 
$\sum_k M_k  = \sum_i m$ and $ \sum_k M_k {\bf U}_k = \sum_i m {\bf v}_i$,
so that if mass, momentum and energy are conserved at the MD 
level, they are also conserved at the DP level.

In order to obtain the equations of motion for the 
DPD we now take the time derivatives  of  Eqs.~(\ref{DPdef}).
The Gaussian form of $s$  makes it possible to write
the time-derivative 
$\dot{f}_k ({\bf x}_i ) = f_{kl}({\bf x}_i ) ({\bf v}'_i \cdot {\bf r}_{kl}
+ {\bf x}'_i \cdot {\bf U}_{kl} )$
where the  function $f_{kl}$ is defined as
$f_{kl} ({\bf x} )\equiv (2/a^2 )f_{k} ({\bf x} ) f_{l} ({\bf x} ) $.
After some algebra \cite{flekkoy98f}
the microscopic equations of motion  then take the form 
\begin{eqnarray}
\frac{\text{d} {M_k}}{\text{d} t} & = & \sum_l \dot{M}_{kl}  \equiv \sum_i  {f}_{kl} ({\bf x}_i )m (
   {\bf v}_i'\cdot {\bf r}_{kl} + {\bf x}'_i \cdot {\bf U}_{kl}   )
\nonumber \\
\frac{\text{d}  {\bf P}_k }{\text{d} t} &=&  M_k {\bf g} +
\sum_l  \dot{M}_{kl} \frac{ {\bf U}_{k}+{\bf U}_{l}}{2} +       \sum_{li}  f_{kl}({\bf x}_i ){\bf \Pi}_i'  
 \cdot {\bf r}_{kl}  \nonumber \\
\frac{\text{d} E_k}{\text{d} t} &=&
\sum_l \frac{ \dot{M}_{kl}}{2} \left( \frac{{\bf U}_{kl}}{2} \right)^2
+ \sum_{li} f_{kl} ({\bf x}_i)\left(  {\bf J}_{ i}'
- \Pi'_i \cdot \frac{{\bf U}_{kl}}{2} \right)
\cdot {\bf r}_{kl}  \label{micro}
\end{eqnarray}
where we have defined
the general momentum-flux tensor
${\bf \Pi}_i' = m {\bf v}_i' {\bf v}_i' + (1/2) \sum_{j}
{\bf F}_{ij} \Delta {\bf x}_{ij} \label{momentum_flux} $, where 
${\bf F}_{ij}$ is the force between MD particles $i$ and $j$,
and the microscopic energy flux vector
$ {\bf J}_{ i}' = \epsilon_i {\bf v}_i' + (1/4)  \sum_{i\neq j}
{\bf F}_{ij} \cdot ({\bf v}_i' + {\bf v}_j') \Delta {\bf x}_{ij} $.
We have also  used the definitions 
 ${\bf v}'_i = {\bf v}_i -  ({\bf U}_k + {\bf U}_l)/2$, ${\bf x}'_i = {\bf x}_i -  ({\bf r}_k + {\bf r}_l)/2$, ${\bf U}_{kl} = {\bf U}_k - {\bf U}_l$, ${\bf r}_{kl} = {\bf r}_k - {\bf r}_l$ and $m{\bf g} $ is the external force 
on an MD particle. 
In the mass equation above the ${\bf x}'_i \cdot {\bf U}_{kl}$ term may be shown 
to be negligible upon averaging, as it samples the difference in mass density 
rather than the average of this quantity 
across the region where $f_{kl} \neq 0$ \cite{flekkoy98f}.
For that reason it will be omitted, and we have already omitted the corresponding terms
in the  momentum  and energy equations.

\begin{figure} 
\centerline{\hbox{\psfig{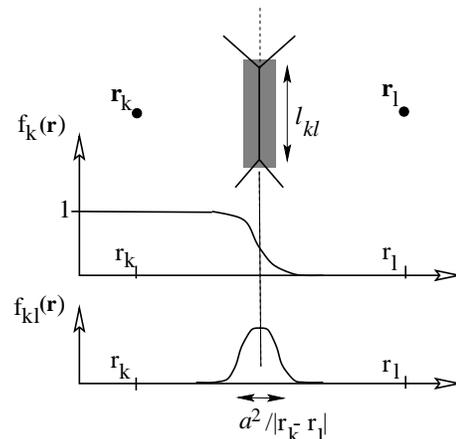}}}
\caption{\label{fig2}
\protect \narrowtext The overlap region between two Voronoi 
cells is shown in grey. The sampling function $f_k ({\bf r} )$ is shown in the
upper graph and the overlap function in the lower graph. 
The width of the overlap
region is $ a^2 / |{\bf r}_k - {\bf r}_l |$ and its length is denoted by $l_{kl}$.}
\end{figure}

All the interaction terms in the above transport equations 
are weighted by  the overlap function $f_{kl} ( {\bf x} ) $.
If only two DP's, $k$ and $l$ say, are present it may be shown that 
$f_{kl} ({\bf x} )=
({1}/(2a^2)) \cosh^{-2}{(({\bf x} - ({\bf r}_k + {\bf r}_l)/2) \cdot ({\bf r}_k - {\bf r}_l)/a^2 )}$.
This function becomes exponentially small away from the dividing line that 
is equally far from ${\bf r}_k$ and ${\bf r}_l$, as is illustrated in Fig.~\ref{fig2}.  
The set of all such dividing lines defines a Voronoi lattice.
In Fig.~\ref{fig1} fictitious MD particles are plotted where $f_{kl}({\bf x} )> 0.2 a^2$.
This happens in the neighborhood of the dividing lines.
When additional  DP's are present their  contribution to the
$\cosh^{-2}$ result may be shown to be negligible in the vicinity 
of the dividing line, except at the corners, where dividing lines meet.
In the end the DPD equations of motion will turn out to
be independent of $a$, and only the length $l_{kl}$ shown in
Fig.~\ref{fig2} will remain. At this point it suffices to construct the Voronoi lattice
itself, and there is no need to evaluate the overlap functions.
Standardized algorithms and software for the construction of Voronoi 
lattices exist \cite{guibas92}.

Note that since the right hand side terms of
the mass and momentum part of Eqs.~(\ref{micro}) are all odd under the 
exchange $l\leftrightarrow k$
the DP's interact in a pairwise and explicitly conservative fashion.
The same is true for the energy equation if it is rewritten
in terms of the total, rather than the internal energy.


Splitting Eqs.~(\ref{micro}) into  fluctuating and average parts
 gives 
\begin{eqnarray}
\frac{\text{d} {M_k}}{\text{d} t} & = &  \sum_{li}  {f}_{kl} ({\bf x}_i )m 
   \langle {\bf v}_i' \rangle \cdot {\bf r}_{kl}   +  \sum_l \dot{\tilde{M}}_{kl} \nonumber \\
  \frac{\text{d}  {\bf P}_k  }{\text{d} t} &=&   M_k  {\bf g}   + \sum_l      \left( \sum_i  f_{kl}({\bf x}_i )
\langle  {\bf \Pi}^i \rangle   \cdot {\bf r}_{kl} \right) + \tilde{\bf F}_{kl} \nonumber \\
\frac{\text{d} E_k}{\text{d} t}  &=& \sum_{li} f_{kl} ({\bf x}_i)\left(  \langle {\bf J}_{ i}' \rangle
- \langle \Pi'_i \rangle \cdot \frac{{\bf U}_{kl}}{2} \right)
\cdot {\bf r}_{kl} \nonumber \\
&+& \sum_l \tilde{\bf F}_{kl} \cdot \frac{{\bf U}_{kl}}{2} + \tilde{q}_{kl}\; . 
\label{dpd1} 
\end{eqnarray}
where $\dot{\tilde{M}}_{kl}$ is the fluctuating part
of the mass flux,   $\tilde{F}_{kl} $ is the fluctuating part of the momentum flux 
$\sum_i f_{kl}({\bf x}_i )  {\bf \Pi}^i \cdot {\bf r}_{kl} +
\dot{{M}}_{kl}({\bf U}_k + {\bf U}_l)/2$,  and $\tilde{q}_{kl}$ 
the fluctuating part of the energy  flux 
$\sum_i f_{kl}({\bf x}_i ) {\bf J}_{ i}' \cdot 
{\bf r}_{kl} + (1/2) \dot{{M}}_{kl} ({\bf U}_{kl}/2)^2 $.
Note that we  have absorbed the contributions 
from  the mass variations in $\tilde{\bf F}_{kl}$ and $\tilde{q}_{kl}$.
The thermal averages,  $\langle ... \rangle$, are 
computed by means of an ensemble
of systems with common {\it instantaneous} values of the mesoscopic
variables $\{ {\bf r}_k , M_k ,  {\bf U}_k, E_k \}$.
This means that only the time derivatives of this set 
have a fluctuating part. 

It is necessary to introduce {\em some} average
description of the MD system.
For this purpose we assume the  
scale separation  $a << {|{\bf r}_k - {\bf r}_l|}$, 
for all $k$ and $l$, and that the width of the overlap region 
$a^2/r_{kl}$ is larger than the mean free path of the MD particles.
For simplicity  we choose the  momentum 
flux tensor of a simple Newtonian fluid
which has  the form  $
\langle {\bf \Pi }^i_{kl}\rangle  = {\bf I} P - \eta (\nabla {\bf v} + (\nabla {\bf v} )^T)$
where $\eta$ is the dynamic viscosity and $P$ the pressure
of the MD fluid, $^T$ denotes the transpose and $\bf I$ is the
identity tensor\cite{landau59}.
We shall make  the approximation that 
the average molecular particle velocity 
$\langle {\bf v} \rangle$ interpolates linearly between the DP's.

It follows that the average 
mass current $\langle {\bf v}' \rangle = {\bf 0 }$
and that the velocity gradients in the momentum-flux tensor 
take the form 
$\nabla {\bf v} + (\nabla {\bf v}  )^T = {1}/{r_{kl}} 
( {\bf e}_{kl} {\bf U}_{kl}  + {\bf U}_{kl} {\bf e}_{kl}  )$, where
${\bf e}_{kl} = ({\bf r}_{k}-{\bf r}_l)/|{\bf r}_{k}-{\bf r}_l|$.
Since $\langle {\bf v}' \rangle \approx {\bf 0}$ we may choose 
the heat flux according to Fouriers law $\langle {\bf J}' \rangle =
\lambda \nabla T$, where 
$T$ is the temperature and $\lambda $ is 
the thermal conductivity. In other words, in the frame of reference of
the overlap region the energy flux is simply the heat flux since 
work terms proportional to the velocity vanish
\cite{landau59}. With this input we get 
$\dot{ {M_k}} = \sum_l \dot{\tilde{M}}_{kl}$ and 
\begin{eqnarray}
  \frac{\text{d}   {\bf P}_k }{\text{d} t} & = &
 M_k  {\bf g} -  \sum_l   l_{kl}  \left( 
\frac{p_{kl}}{2} {\bf e}_{kl} +\frac{\eta}{r_{kl}}  \left( {\bf U}_{kl}  + ({\bf U}_{kl}
  \cdot  {\bf e}_{kl})  {\bf e}_{kl}  \right)
\right)\nonumber \\
&+& \sum_l \tilde{\bf F}_{kl} \nonumber \\
\frac{\text{d} E_k}{\text{d} t}   &=& \sum_l l_{lk}  \left( \frac{p_{k}+p_l}{2} {\bf e}_{kl} + \frac{\eta}{r_{kl}}
({\bf U}_{kl} + ({\bf U}_{kl}\cdot {\bf e}_{kl}){\bf e}_{kl}) \right)  \cdot
\frac{{\bf U}_{kl}}{2}  \nonumber \\
&+&  \sum_l l_{lk}   \lambda \frac{T_{kl} }{r_{kl}} + \tilde{\bf F}_{kl} \cdot \frac{{\bf U}_{kl}}{2} + \tilde{q}_{kl}\; .
\label{dpd2} 
\end{eqnarray}
where we have assumed that the pressure $p$ and temperature $T$, as 
well as the average velocity, interpolates linearly between DP centers, 
 $p_{kl} = p_k - p_l$ and   $T_{kl} = T_k - T_l$.
The pressure will eventually follow from an equation of state
of the form $p_k= p(E_k,V_k)$ where $V_k$ is the volume 
of DP $k$.
The pressure and temperature must be obtained via 
a thermodynamic description, i.e. an equation 
of state that relates pressure and temperature to energy $E_k$
and volume $V_k$.
In the special case of an ideal gas (see Ref. \cite{flekkoy98f}
for a more general treatment), these relations simplify to
$\text{d} p_k =p_k ({\text{d} E_k}/{E_k} - {\text{d} V_k}/{V_k} )$ and
$\text{d} T_k =  \text{d} E_k /({Nk_B})$.

The average rate of change of $M_k$ vanishes.
This  allows us to neglect mass variations  
altogether in the DPD equations,
since the effect of mass fluctuations may then be absorbed  in 
$\tilde{\bf F}_{kl}$ and $\tilde{q}_{kl}$. Had there been a 
coupling, say, between the {\it averaged} momentum and mass values
the mass would have had to be updated as well.
With nothing but fluctuations in $M_k$ 
the only change  introduced by  the $M_k=$const. approximation is 
the loss of the drift in the $M_k$'s around their constant average, caused by 
the fluctuations. Nothing is neglected in the instantaneous changes
of momentum and energy.

In general, force fluctuations will
cause mass fluctuations, which in turn will couple back to 
cause momentum fluctuations. The time scale over which this will
happen is $t_{\eta}= r_{kl}^2/\eta$, where 
$\eta$ is the dynamic viscosity of the MD system.
This is the time it takes for 
a velocity perturbation to decay over a distance $r_{kl}$.
We shall need to make the
assumption that the fluctuating forces are Markovian, 
and it is clear that 
this assumption may only be valid on time scales larger than $t_{\eta}$.
Since the time scale of a hydrodynamic perturbation of size $l$, say, is 
also given as $l^2/\eta$ this restriction implies  the scale separation
requirement $r_{kl}^2 << l^2$, consistent with the scale $r_{kl}$ being
mesoscopic.

In `conventional' DPD 
\cite{hoogerbrugge92,marsh98c,espanol95b}
the forces are pairwise and act parallel to ${\bf e}_{kl}$. They 
have a conservative part that depends only on $r_{kl}$
and a dissipative part proportional to 
$({\bf U}_{kl}\cdot {\bf e}_{kl}){\bf e}_{kl}$. 
Here the same terms are present. The conservative
force is seen to arise  from the pressure
and the dissipative part from dissipation in the underlying fluid
with the MD viscosity taking the place of a 
postulated friction  coefficient.
In addition there is a dissipative term parallel to ${\bf U}_{kl}$.
The energy part of Eq.~(\ref{dpd2}) 
is identical in form to the energy equation postulated 
by Avalos and Mackie  \cite{avalos97}
and similar to the equation studied by Espa\~{n}ol
\cite{espanol97}, save for the fact that here the work done by the  conservative 
force $ ({p_{k}+p_l}) {\bf e}_{kl} \cdot {{\bf U}_{kl}}/{4}$ is present.  
The principal difference is associated with the Voronoi lattice:
Our DP's fill space and change their shape, a key feature
that enables this new DPD to treat a multitude of length scales within
a single simulation.

In order to obtain $\tilde{{\bf F}}$ and $\tilde{q}$ 
we invoke the Markovian approximation to write
$ \tilde{\bf F} = \mbox{\boldmath$\omega$}_{kl \parallel} W_{kl\parallel}  + \mbox{\boldmath$\omega$}_{kl \perp} 
W_{kl\perp}$, 
where ${\bf \tilde{F}}_{kl}$ is decomposed into components parallel
and perpendicular to ${\bf e}_{kl}$, and 
the $W$'s are defined as Gaussian random variables 
with the  correlation function 
$\langle  W_{kl\alpha } (t) W_{nm \beta }(t') \rangle 
= \delta_{\alpha \beta} \delta (t-t')   (\delta_{kn}\delta_{lm}
+\delta_{km}\delta_{ln}) $
where $\alpha $ and $\beta $ denotes either $\perp$ or $\parallel$.
Newton's third law  
guarantees that $\mbox{\boldmath$\omega$}_{kl} 
= - \mbox{\boldmath$\omega$}_{lk}$. 
We have the similar expression $\tilde{q}_{kl} = \Lambda_{kl} W_{kl}$
where $\Lambda_{kl}  = -\Lambda_{lk} $ and $W_{kl}$
satisfies the above equation for the $W$'s without the 
$\delta_{\alpha \beta}$-factor.

In Refs. \cite{espanol95,flekkoy98f,avalos97}
the magnitudes of $\tilde{\bf F}_{kl}$ 
and $\tilde{q}_{kl}$ are obtained
on the basis of the Fokker-Planck equation 
which derives from equations like Eqs.~(\ref{dpd2}).
The  isothermal results, adapted to the present case,
are the fluctuation dissipation relations
$ \omega_{kl\parallel}^2 = 2 \omega_{kl\perp}^2 
= 4 \eta k_BT({l_{kl}}/{r_{kl}}) $
for the force $\tilde{\bf F}_{kl}$  and for the heat fluctuations 
 $\Lambda_{kl}^2 = 2 k_B T \lambda ({l_{kl}}/{r_{kl}})$ \cite{avalos97}.
It is also possible to show that detailed balance \cite{gardiner85}
holds, and that the DP's obey the Gibbs distribution \cite{flekkoy98f}
$ \rho^{\text{eq}} =  Z^{-1}(T,V) 
 \exp{\left( -\beta \sum_k \left( {P_k^2}/{2M_k}
+ V({\bf r}_k ) \right) \right) }$, where the potential $V({\bf r}_k )$ is responsible for the 
pressure force in Eq.~(\ref{dpd2}), and $T=1/(\beta k_B)$
is the temperature characterizing the MD system~\cite{flekkoy98f}.

Considering now the application illustrated in Fig.~\ref{fig1}
we need to define DP--colloid forces.
Taking  the hydrodynamic momentum flux tensor and 
Eq.~(\ref{dpd2}) as a starting point we observe that 
the  DP--colloid interaction may be 
obtained in the same form as the DP--DP interaction 
of Eq.~(\ref{dpd2}) with the replacement of $l_{kl}\rightarrow L_{kI}$,
where  $L_{kI}$  is the length (area in 3D) of the arc segment where the DP
meets the colloid (see Fig.~\ref{fig1}). The  velocity 
gradient is that between the DP 
and the colloid surface. The latter may be computed 
by linear interpolation 
using ${\bf U}_k$ and the velocity of the colloid surface 
together with a no-slip boundary condition on this surface.
In the momentum fluctuation-dissipation relation 
too the replacement $l_{kl}\rightarrow L_{kI}$
must be made. In order to increase the spatial resolution where 
colloidal particles are close it is necessary to introduce a higher 
DP density there; this ensures that fluid lubrication effects are 
maintained.
After these particles have moved it may be necessary to re-tile
the DP system, as is done in finite element 
calculations. This is simply achieved by distributing 
the mass, momentum and energy  of the old DP's on the new ones according to 
their area (or volume in 3D).
When a new Voronoi cell is created on top of the 
old ones the new DP mass is simply the 
sum of the old DP mass densities times
the area of overlap between the new and old dissipative particles 
(and similarly
for the DP momentum and energy).
This technique may also be used to model polymer 
solutions if the polymer chains are formed of linked beads.

The DPD which we have derived in the present work
is similar to conventional DPD, without-~\cite{espanol95}
and with energy conservation \cite{avalos97} . 
But the forces conventionally used to define DPD have now been 
given a microscopic basis. More important, however, is the fact that our 
analysis permits the introduction of specific
physical interactions at the mesoscopic level, together with
a well-defined meaning for this mesoscale.
Finally, we note the similarity of the present
particulate description, which is based on a bottom-up approach, 
to existing continuum approaches, which start out from a 
macroscopic description. Such top-down approaches include in particular 
smoothed particle hydrodynamics~\cite{monaghan92} and 
finite-element simulations.
In these descriptions too the computational method is based 
on tracing the motions of elements of the fluid on the basis of the forces 
acting between them~\cite{boghosian98}.  
We stress, however, that while such top-down computational 
strategies require as initial input a 
macroscopic phenomenological description,
the present approach relies on a microscopic 
representation from the outset.


\begin{thebibliography}{10}

\bibitem{frisch86}
U. Frisch, B. Hasslacher, and Y. Pomeau, Phys. Rev. Lett. {\bf 56},  1505
  (1986).

\bibitem{koplik95}
J. Koplik and J.~R. Banavar, Ann. Rev. Fluid Mech. {\bf 27},  257  (1995).

\bibitem{hoogerbrugge92}
P.~J. Hoogerbrugge and J.~M.~V.~A. Koelman, Europhys. Lett. {\bf 19},  155
  (1992).

\bibitem{boek96}
E.~S. Boek, P.~V. Coveney, H.~N.~W. Lekkerkerker, and P. van~der Schoot, Phys.
  Rev. E {\bf 54},  5143  (1996).

\bibitem{schlijper95}
A.~G. Schlijper, P.~J. Hoogerbrugge, and C.~W. Manke, J. Rheol. {\bf 39},  567
  (1995).

\bibitem{coveney96}
P.~V. Coveney and K.~E. Novik, Phys. Rev. E {\bf 54},  5143  (1996).

\bibitem{groot97}
R.~D. Groot and P.~B. Warren., J. Chem. Phys. {\bf 107},  4423  (1997).

\bibitem{espanol95}
P. Espanol, Phys. Rev. E {\bf 52},  1734  (1995).

\bibitem{marsh98c}
C.~A. Marsh and P.~V. Coveney, J. Phys. A: Math. Gen. {\bf 31},  6561  (1998).

\bibitem{hole88}
K. Ho-Le, Computer Aided Design {\bf 20},  27  (1998).

\bibitem{coveney98}
P.~V. Coveney {\it et~al.}, Int. J. Mod. Phys. C {\bf 9}, 1479 (1998).

\bibitem{flekkoy98f}
E.~G. Flekk{\o}y and P.~V. Coveney, preprint (unpublished).

\bibitem{guibas92}
D.~E.~K. L.~J.~Guibas and M. Sharir, Algorithmica {\bf 7},  381  (1992).

\bibitem{landau59}
L.~D. Landau and E.~M. Lifshitz, {\em Fluid Mechanics} (Pergamon Press, New
  York, 1959).

\bibitem{espanol95b}
P. Espanol and P. Warren, Europhys. Lett. {\bf 30},  191  (1995).

\bibitem{avalos97}
J.~B. Avalos and A.~D. Mackie, Europhys. Lett. {\bf 40},  141  (1997).

\bibitem{espanol97}
P. Espanol, Europhys. Lett {\bf 40},  631  (1997).

\bibitem{gardiner85}
C.~W. Gardiner, {\em Handbook of Stochastic Methods} (Springer Verlag, Berlin
  Heidelberg, 1985).

\bibitem{monaghan92}
J.~J. Monaghan, Ann. Rev. Astron. Astophys. {\bf 30},  543  (1992).

\bibitem{boghosian98}
B.~M. Boghosian, Encyclopedia of Applied Physics {\bf 23},  151  (1998).

\end{thebibliography}

\end{multicols}
\end{document}